\pdfoutput=1
\documentclass[reprint,aps,prl,floatfix,twocolumn,superscriptaddress,showpacs,amsmath,amssymb,showpacs]{revtex4-1}
\usepackage{graphicx}
\usepackage{epstopdf}
\usepackage{epsfig}
\usepackage{epsf}
\usepackage{url}
\usepackage[USenglish]{babel}
\usepackage{hyperref}
\def\bcen{\begin{center}}
\def\ecen{\end{center}}
\allowdisplaybreaks
\renewcommand\[{\begin{equation}}
\renewcommand\]{\end{equation}}
\usepackage{verbatim}
\usepackage{xcolor}
\usepackage{natbib}

\begin{document}
\title{Hund excitations and the efficiency of Mott solar cells}
\author{Francesco Petocchi}
\affiliation{Department of Physics, University of Fribourg, 1700 Fribourg, Switzerland}
\author{Sophie Beck}
\affiliation{Materials Theory, ETH Zurich, Wolfgang-Pauli-Strasse 27, 8093 Z\"urich, Switzerland}
\author{Claude Ederer}
\affiliation{Materials Theory, ETH Zurich, Wolfgang-Pauli-Strasse 27, 8093 Z\"urich, Switzerland}
\author{Philipp Werner}
\affiliation{Department of Physics, University of Fribourg, 1700 Fribourg, Switzerland}
\begin{abstract}
We study the dynamics of photo-induced charge carriers in realistic models of LaVO$_3$ and YTiO$_3$ polar heterostructures. It is shown that two types of impact ionization processes contribute to the carrier multiplication in these strongly correlated multi-orbital systems: The first mechanism involves local spin state transitions, while the second mechanism involves the scattering of high kinetic energy carriers. Both processes act on the 10 fs timescale and play an important role in the harvesting of high energy photons in solar cell applications. As a consequence, the optimal gap size for Mott solar cells is substantially smaller than for semiconductor devices. 
\end{abstract}


\maketitle

\paragraph*{Introduction -} The development of efficient photovoltaic technologies is essential for a sustainable energy production. Over the last decade, the emergence of perovskite solar cells \cite{Kojima2009,Im2011} has caught the attention of many researchers. The efficiency of metal halide perovskite devices has risen rapidly, but practical issues like stability and toxicity still need to be solved \cite{Assadi2018}. Much less in the spotlight, but conceptually interesting, is the proposal to build solar cells which expoit the strongly interacting nature of electrons in Mott insulators. In Ref.~\onlinecite{Manousakis2010} it was pointed out that charge carriers with high kinetic energy in small-gap Mott insulators can excite additional electrons across the gap via impact ionization. This provides a mechanism for harvesting high-energy photons, which may potentially lift the efficiency of Mott insulating solar cells above the Shockley-Queisser limit \cite{Shockley1961} for semiconductor solar cells. A promising system, LaVO$_3$ (LVO) on top of SrTiO$_3$ (STO), was identified by Assmann {\it et al.} in Ref.~\onlinecite{Assmann2013}. It has a direct band-gap of about 1.1 eV and an internal electric field due to the polar nature of the LVO layers. Experimentally, however, the devices fabricated so far have shown a low efficiency due to a low mobility of the photo-carriers \cite{Wang2015,Jellite2018}. Also theoretically, it has been argued that the strong internal fields of these structures may localize charge carriers \cite{Eckstein2014}. To properly assess the device characteristics of clean Mott insulating heterostructures, it is essential to analyze the impact ionization processes in realistic multi-orbital systems and the charge separation process in the presence of an external voltage bias.

\paragraph*{Method -}
We use density-functional theory (DFT) to calculate the band structures for bulk LVO, as well as for multilayers composed of four LVO layers and two STO layers, respectively, thereby modelling semi-infinite STO substrates on both sides of the film. Within an energy window containing the predominantly $t_{2g}$-derived bands around the Fermi level, we construct three $t_{2g}$-like maximally localized Wannier functions (MLWFs) \cite{Marzari_et_al:2012} centered on each transition metal site to obtain an effective tight-binding Hamiltonian $\mathcal{H}_{0}$ for the system. The local interaction is described by the three-orbital Slater-Kanamori Hamiltonian
%
\begin{eqnarray}
\mathcal{H}_{U} & = & U\underset{\alpha}{\sum}\hat{n}_{\alpha\uparrow}\hat{n}_{\alpha\downarrow}+
U'\underset{\alpha\neq\beta}{\sum}\hat{n}_{\alpha\uparrow}\hat{n}_{\beta\downarrow}+
U''\underset{\sigma,\alpha<\beta}{\sum}\hat{n}_{\alpha\sigma}\hat{n}_{\beta\sigma}\nonumber\\
 & + & J_{H}\underset{\alpha\neq\beta}{\sum}c_{\alpha\uparrow}^{\dagger}c_{\beta\uparrow}
 c_{\beta\downarrow}^{\dagger}c_{\alpha\downarrow}+J_{H}\underset{\alpha\neq\beta}{\sum}
 c_{\alpha\uparrow}^{\dagger}c_{\beta\uparrow}c_{\alpha\downarrow}^{\dagger}c_{\beta\downarrow},
 \label{ham}
\end{eqnarray}
%
where $c_{\alpha\sigma}$ $(c_{\alpha\sigma}^{\dagger})$ is the annihilation (creation) operator of an electron of spin $\sigma$ in orbital $\alpha$ and the interaction parameters are the local intra(inter)-orbital Coulomb 
interaction $U$ ($U'$) and Hund's coupling $J_{H}$. We choose $U'=U-2J_{H}$, $U=4.5$ eV and $J_H=0.64$ eV  in accordance with previous studies on bulk LVO \cite{DeRaychaudhury2007,Ederer2}. To investigate the photo-induced dynamics in this model, we employ a non-equilibrium implementation of inhomogeneous dynamical mean-field theory (DMFT) \cite{Eckstein2013,Aoki2014}, where the lattice system is mapped onto a coupled set of six impurity problems, one for each layer.
This approach assumes a local self-energy, but retains time-dependent charge fluctuations \cite{Georges1996}. The impurity problem is solved using a non-crossing approximation (NCA) solver \cite{Keiter1971,Eckstein2010}, which provides a qualitatively correct description of Mott insulating non-equilibrium systems \cite{Eckstein2010breakdown,Ligges2018}. All calculations are performed at $T=0.1$ eV.

An inhomogeneous nonequilibrium DMFT simulation with explicit momentum summation and 3 orbitals per site would be limited to very short times. We thus implement a simplified self-consistency which is more economical, both in terms of computational cost and memory requirement. Using the cavity method \cite{Georges1996,Werner2017}, the hybridization function $\hat \Delta$ for site $i$ can be expressed as
\begin{equation}
  \hat \Delta_{i}(t,t') = {\sum_j}\hat{h}_{ij}(t)\hat{G}^{\left[i\right]}_{j}(t, t' )\hat{h}^{*}_{ji}(t' ),
  \label{deltadef}
\end{equation}
where $\hat h$ denotes the hopping amplitude and $\hat{G}^{\left[i\right]}_{j}$ the Green's function at site $j$ in a lattice with site $i$ removed. All quantities are matrices in orbital and spin space. In a Mott insulating system, it is reasonable to replace $\hat{G}^{\left[i\right]}_{j}$ with the local Green's function $\hat{G}_{j}$, which results in a set of Bethe-like self-consistency equations \cite{Georges1996}, with realistic hopping parameters, which we truncate at the second neighbor. 
To avoid the calculation of orbital off-diagonal components of $\hat G$, we rotate the hopping matrix to an orbital-diagonal, or crystal field basis. The advantage is twofold: first, the much smaller number of hybridization functions leads to less artificial correlations in the NCA solution, and second, we only need to solve one impurity problem per layer, since the two Green's functions of the unit cell become identical by symmetry. Figure~\ref{fig1} shows the comparison between the equilibrium DMFT spectral functions $A_j( \omega ) = -\text{Im} G_j^{R} ( \omega ) / \pi $ obtained using the full $\mathbf{k}$-summation over the Brillouin zone and an exact-diagonalization (ED) solver, and those obtained using the simplified self-consistency and NCA solver ({\it without} any adjustments of hoppings or other parameters). Both results agree remarkably well with regard to the size of the Mott gap, local energy, as well as the widths and substructures of the Hubbard bands. This demonstrates that our simplified treatment provides a qualitatively correct description of the LVO/STO system. 
\begin{figure}
  \hspace*{-5mm}
  \includegraphics[width=0.4\textwidth]{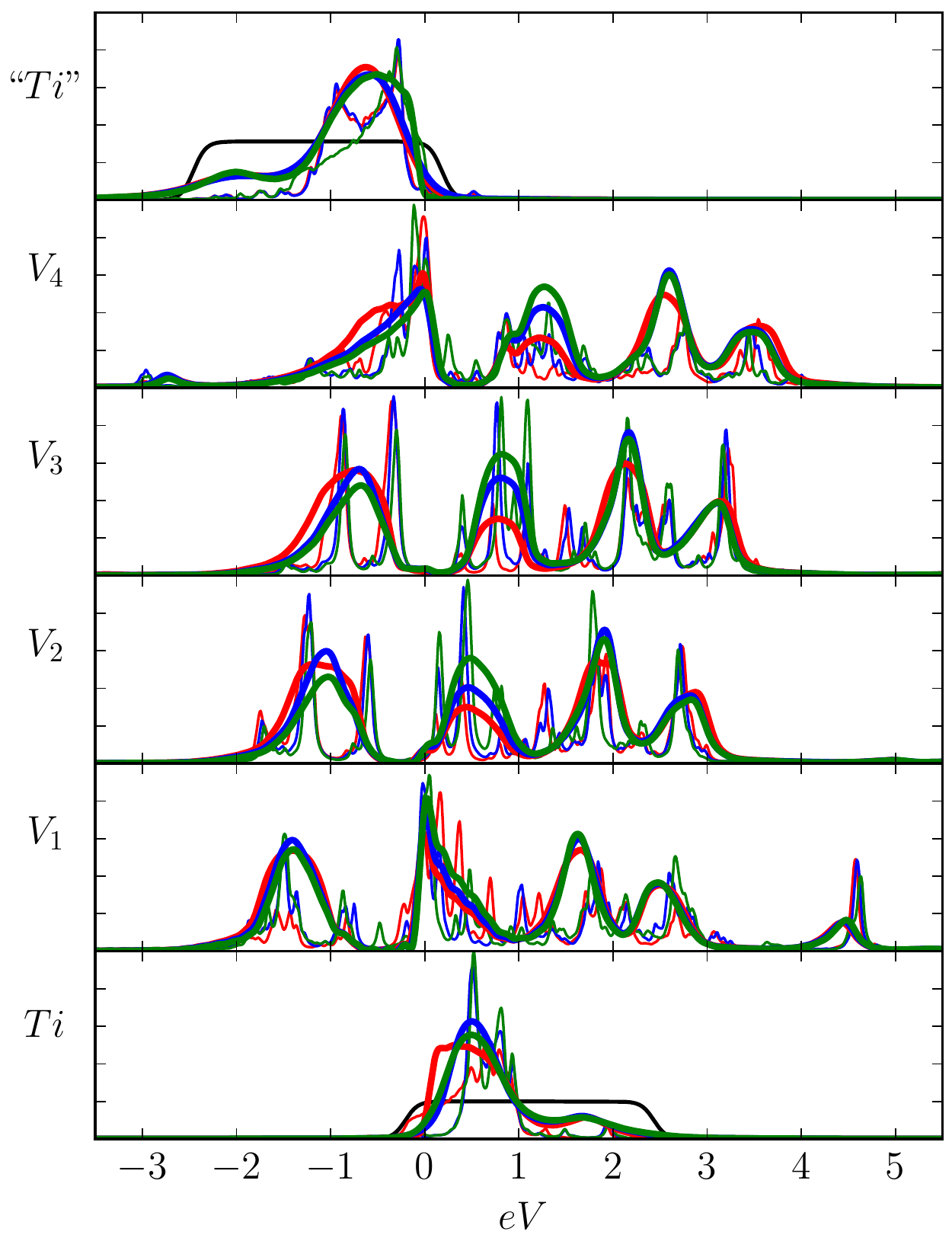}
  \caption{Comparison between the orbital-resolved DMFT spectral functions computed with the full $\mathbf{k}$-summation and an ED solver (thin lines) and those obtained with the approximate self-consistency and NCA solver (thick lines). The black lines indicate the DoS of the metallic leads. The local energy of the uppermost Ti layer has been shifted by hand in order to create a lead DoS representing an electron reservoir. 
\label{fig1}}
\end{figure}
\paragraph*{Light-matter interaction -}
If the multi-orbital basis is not complete, as is usually the case if only a subset of orbitals is considered, the coupling  between the electromagnetic field and matter is in general not gauge invariant \cite{Foreman2002}.  A possible solution is to express the light-matter interaction in terms of the real fields \cite{DenisDP} and to start from the hopping Hamiltonian in the continuous space and dipolar approximation: $\mathcal{H}_{0}+\overrightarrow{E}(t)\cdot\overrightarrow{D}$. Here $\overrightarrow{D}=e\overrightarrow{r}$ is the dipolar matrix that we project onto the Wannier basis. The effect of the light on the system is then given by a time, orbital, and site dependent hopping 
\begin{align}
  h_{\alpha\beta}^{ab}(\mathbf{R}_{i},t)&=e^{i\frac{e}{\hbar}\overrightarrow{A}(t)(\mathbf{R}_{i}+\mathbf{r}_{b}-\mathbf{r}_{a})} \nonumber \\
  &\quad\times\left[h_{\alpha\beta}^{ab}(\mathbf{R}_{i})+\overrightarrow{E}(t)\cdot\overrightarrow{D}{}_{\alpha\beta}^{ab}(\mathbf{R}_{i})\right],
  \label{LMinteract}
\end{align}
with $\vec A(t)=-\int_0^t ds \vec E(s)$ the vector potential and $\alpha$ $(a)$ the orbital (site) index. To mimic a solar-like excitation one could Fourier transform the solar spectrum, but we will scan different photon energies $\Omega$ in order to study the effect of impact ionization. We thus apply in-plane electric field pulses with  Gaussian power-spectra centered on different frequencies, and amplitudes adjusted such that the number of photons is the same for all the pulses \footnote{We used ${ E^2(\Omega,\omega)\sim f_\text{Gauss}(\omega-\Omega)}$ with ${ \int d\omega E^2(\Omega,\omega)/\omega=\text{const}}$ }. In the following analysis we consider the weak field limit in which the dipolar matrix has a minor effect.
\begin{figure}
  \hspace*{-1mm}
  \includegraphics[width=0.228\textwidth]{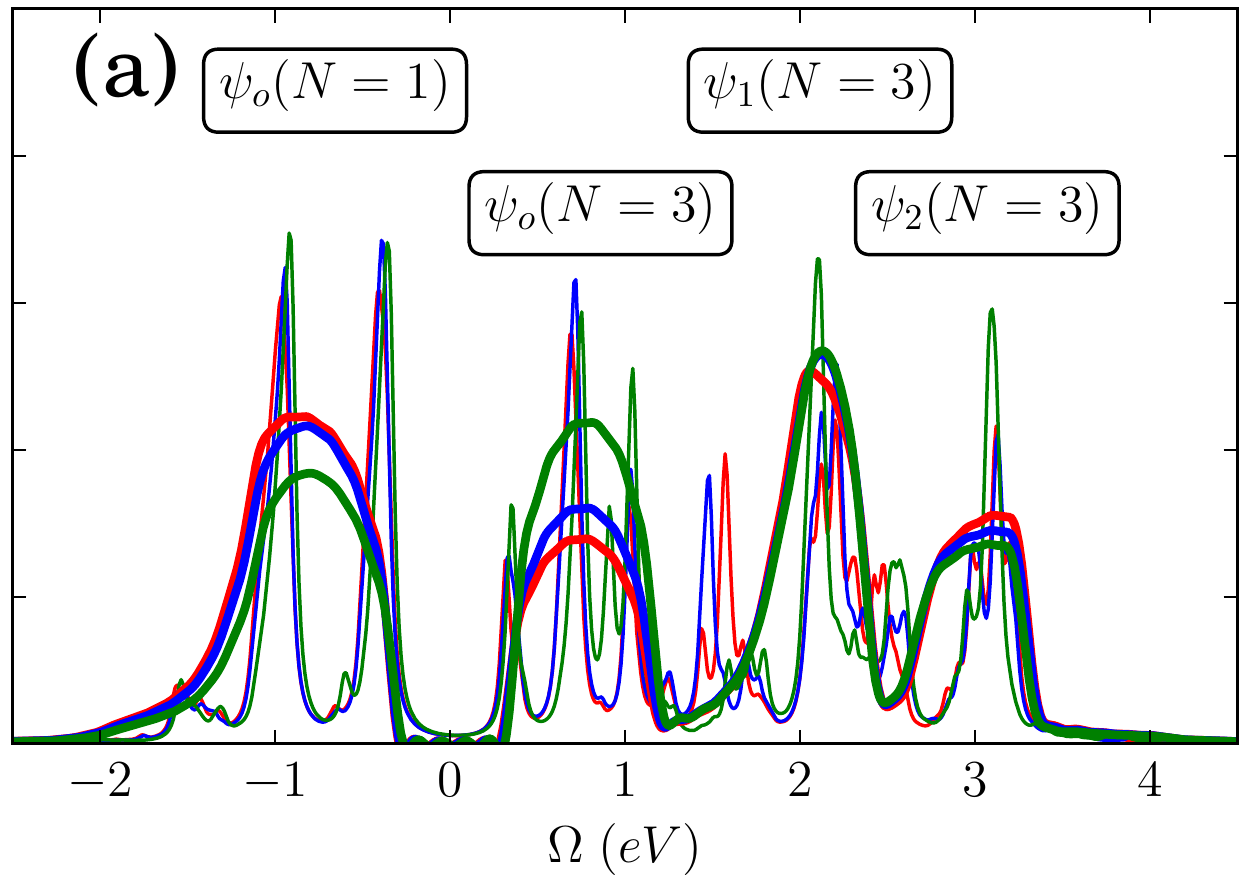}
  \hspace*{1mm}
  \vspace{3mm}
  \includegraphics[width=0.228\textwidth]{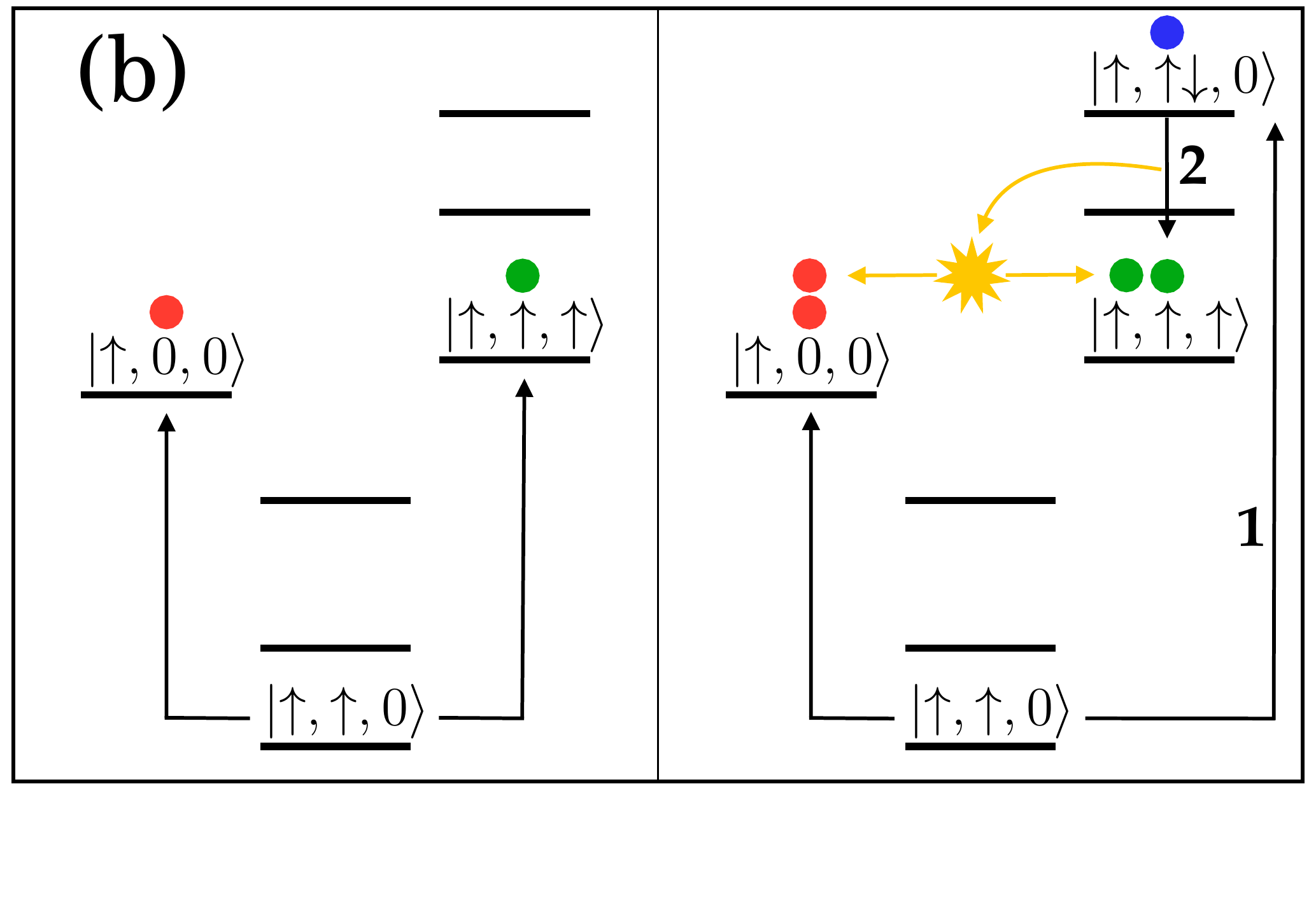}
  \hspace*{-6mm}
   \includegraphics[width=0.29\textwidth]{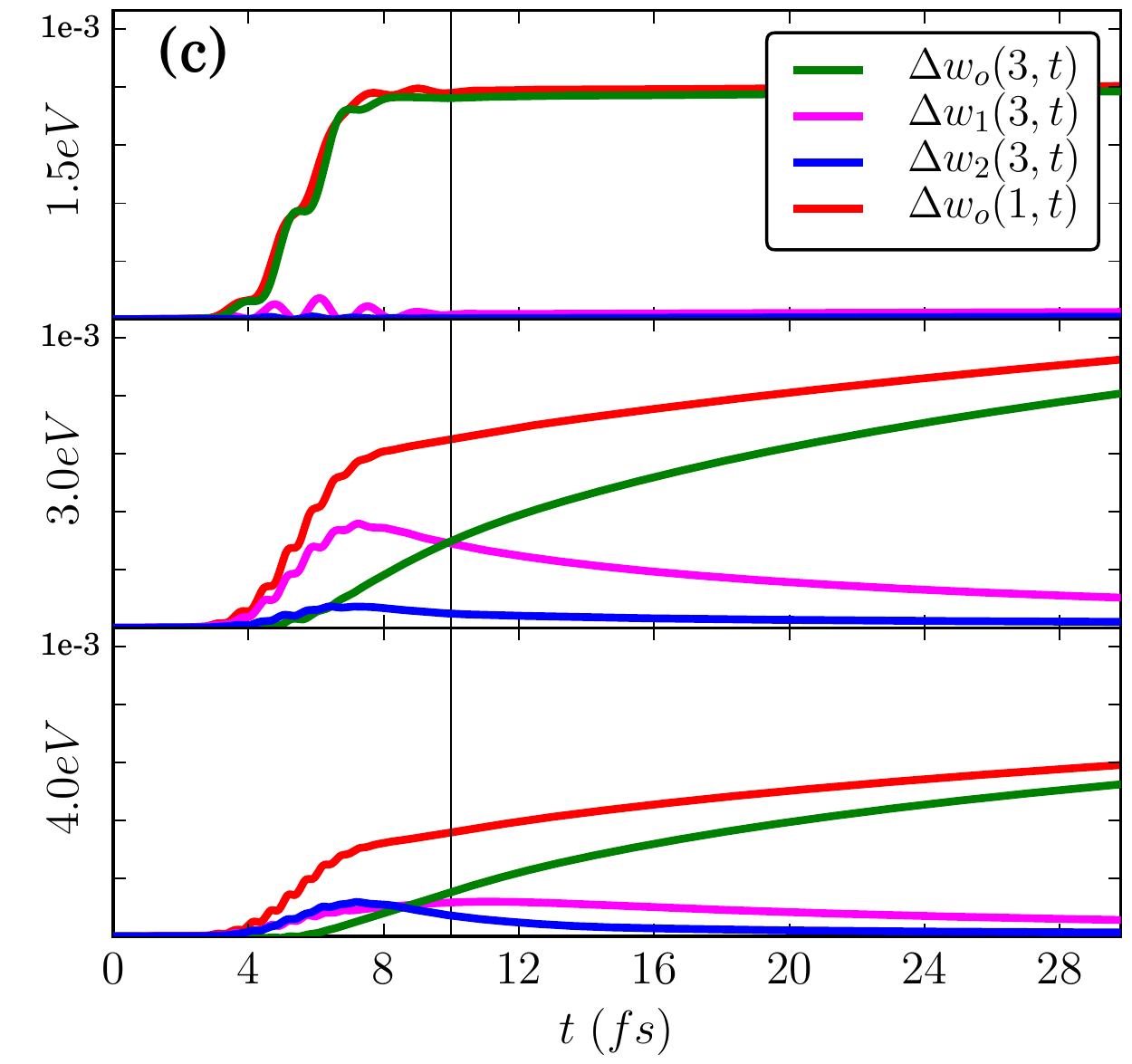}
   \includegraphics[trim=0 -1mm 0 0 , width=0.203\textwidth]{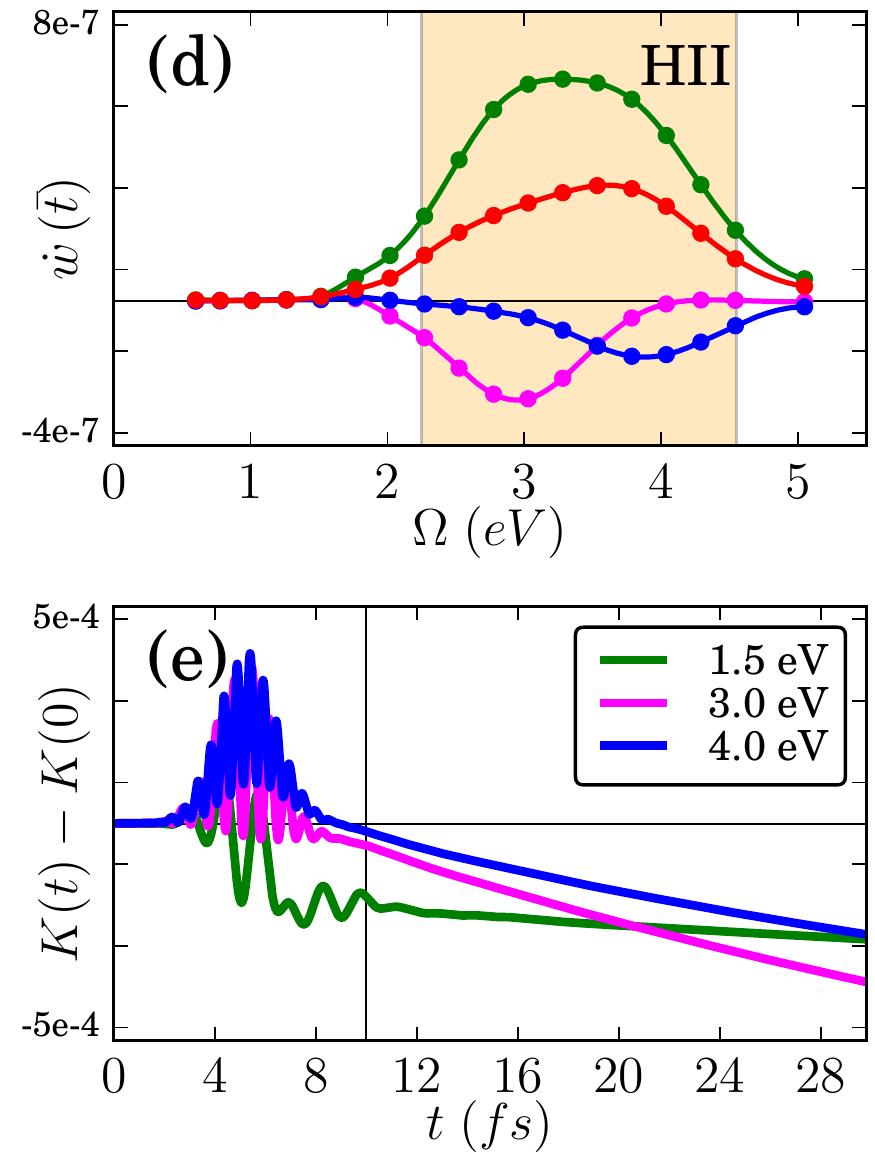}
  \caption{(a) DMFT spectral functions of bulk LVO obtained with the two methods described in the text, with indicated atomic eigenstates. (b) Illustration of the HII process: low energy pulses only produce high spin triplons (left panel), while more energetic pulses initially excite low spin configurations (1), which quickly decay to the high spin state (2). The energy released by the latter process can produce an additional singlon-triplon pair.  (c) Time evolution of the singlon and triplon weights $\Delta w_{n}(N,t)$ for pulses promoting carriers to each of the three triplon subbands. The straight vertical line marks the end of the pulse. (d) Time derivative of $\Delta w_{n}(N,t)$ at $t=12$. Negative values correspond to decaying high energy tiplon states, positive values to accumulating singlon and low-energy triplon states. HII is active in the shaded region. (e) Change in kinetic energy $K$ after the pulse.\label{fig2}}
\end{figure}
\paragraph*{Impact Ionization -}
We start by investigating the carrier multiplication induced by impact ionization \cite{Werner2014} in bulk LVO. For a sector with a given filling $N$, we focus on the $n$-th family of atomic eigenstates characterized by the same $L$ and $S$, collectively denoted by $\left|\psi_{n}(N)\right\rangle$. The corresponding time-dependent contribution $w_n(N,t)$ to the density matrix can be measured directly within NCA. $w_{n}(N,t)$ contains the information on how, during the pulse and the following relaxation, the weights are shifted among the families of atomic eigenstates, resulting in the time dependence of local observables.  In LVO, with two electrons in 3 orbitals, the equilibrium state is dominated by $N=2$ high-spin states, and photo-excitation primarily creates charge carriers in the $N=1$ and $N=3$ sectors (``singlons" and ``triplons"), see Fig.~\ref{fig2}a. Thus an increasing  $\Delta w_{0}(1,t)=w_{0}(1,t)-w_{0}(1,0)$ indicates singlon production, while an increasing $\Delta w_{n}(3,t)$ corresponds to triplon production.  In the following discussion, the index $n$ labels the local spin configurations of the photo-induced states. A pulse with an energy comparable to the gap will create a singlon at the top of the LHB and a triplon in a high-spin configuration at the bottom of the UHB. A more energetic pulse can populate one of the two low-spin configurations of the $N=3$ sector. These states store a considerable amount of {\it Hund energy}, of the order of $5J_{H}$ which can be released by flipping back to the high spin state, and potentially produce additional singlon-triplon pairs. We call this mechanism ``Hund impact ionization" (HII). The second impact ionization channel \cite{Werner2014} involves charge carriers with high kinetic energy scattering from the upper to the lower edge of a Hubbard (sub)band, exciting additional singlon-triplon pairs. This mechanism will be referred to as ``kinetic impact ionization" (KII). Both HII and KII are expected to play a role in the LVO structure, since $5J_H$ and the widths of the $\left|\psi_{n}(N=3)\right\rangle$ subbands are comparable or larger than the gap. In Fig.~\ref{fig2}c we plot $\Delta w_{0}(1,t)$ and $\Delta w_{n}(3,t)$ for photon energies corresponding to excitations from the LHB to the three peaks of the UHB. Exciting triplons into a low-energy high-spin configuration $|\psi_0(3)\rangle$ leads to a long lived state without any carrier multiplication within the accessible time window, as one can infer from the almost constant weights. On the other hand, as soon as one of the two low-spin configurations is populated, the system quickly (even during the pulse) relaxes to the $\left|\psi_{0}(3)\right\rangle$ state. This relaxation leads to a triplon/singlon population that increases in time, consistent with HII. The time derivatives plotted in Fig.~\ref{fig2}d (measured at $t=12$) show that HII is activated for pulse energies $\Omega$ larger than $2$ eV, i.e. as soon as the low-spin states are populated. Since the transition is between two local states, HII does not necessarily result in a lowering of the kinetic energy $K$. Spin state transitions (and the field pulse itself) may produce carriers at the upper edge of the high-spin sub-band $\left|\psi_{0}(3)\right\rangle$. This enables the transformation of kinetic energy into additional singlon-triplon pairs via KII, see Fig.~\ref{fig2}e. After the low energy pulse ($\Omega=1.5$ eV), only KII is active and we observe a rapid decrease of the kinetic energy during and immediately after the pulse, while the HII processes triggered by higher energy pulses repopulate the high-kinetic energy states, which results in a slower decrease of $K(t)$. These results show that the two processes, HII and KII, are intertwined, with HII boosting KII, and that the carrier multiplication by impact ionization occurs on the timescale of a few 10 fs. 
\begin{figure}
  \hspace*{-5mm}
  \includegraphics[width=0.4\textwidth]{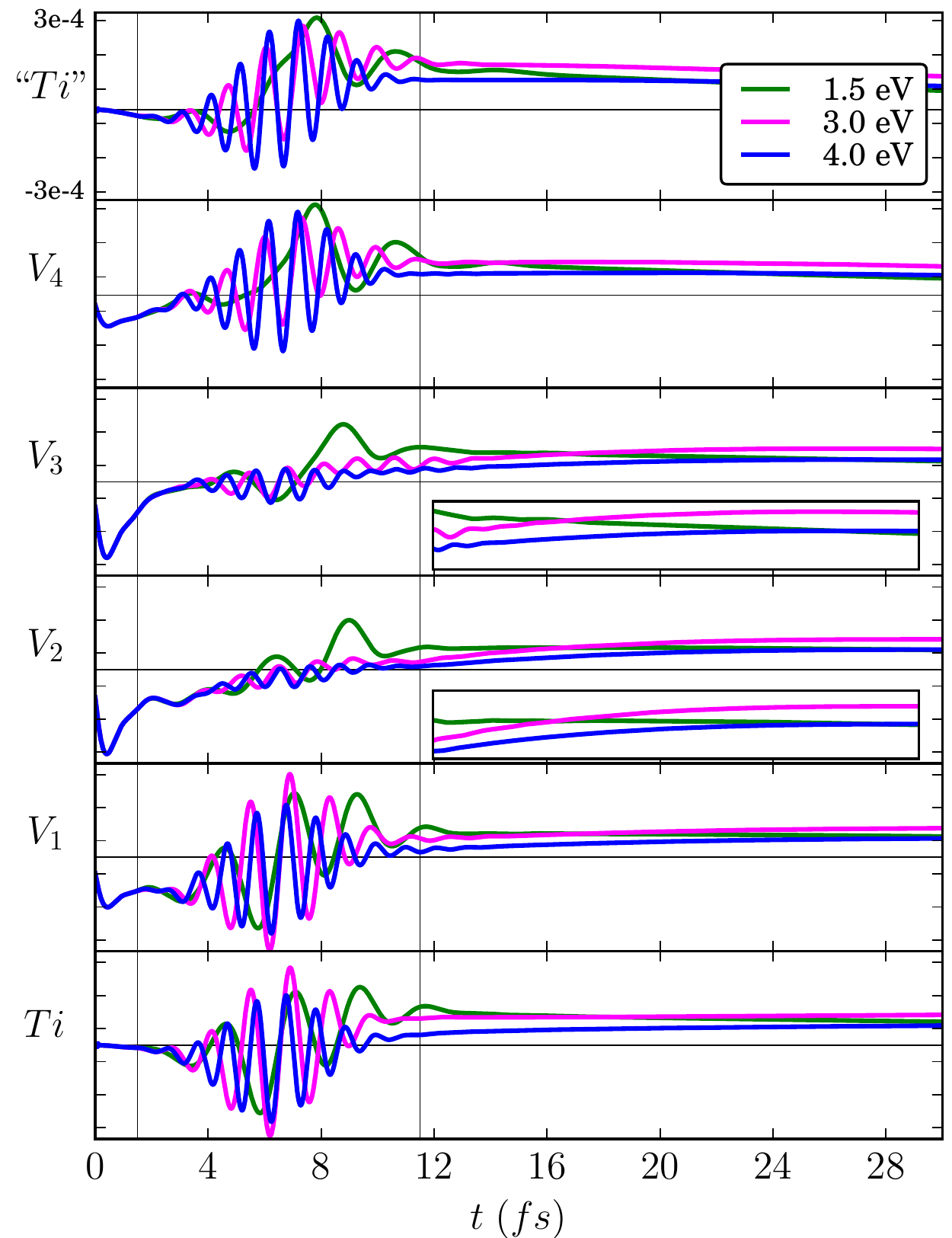}
  \caption{Particle current density from the top to the bottom in the LVO/STO heterostructure of Fig.~\ref{fig1} with applied $V_\text{bias}=0.05$ eV. The initial negative peak is due to the charge rearrangement induced by the bias. The two vertical lines indicate the beginning and end of the pulse. The insets show the increase of the inter-layer current after the pulse for pulse energies compatible with impact ionization.\label{fig3}}
\end{figure}
\paragraph*{Carrier separation -}
We now turn to the LVO/STO heterostructure with internal field gradient and study how the photo-excited charge carriers are transported to the top and bottom leads (Fig.~\ref{fig1}). If the leads, represented in our system by three non-interacting Ti-like $t_{2g}$ bands and a noninteracting electron bath, are at the same chemical potential, all the kinetic energy will be dissipated in the bulk of the leads, which we mimic by a flat electronic density of states (DoS). To harvest energy, it is necessary to apply an external bias $V_\text{bias}$, which counteracts the internal field gradient, in order to collect the electron and hole-like carriers at different chemical potentials. In the simulations, starting from an equilibrium state without external bias, we smoothly switch on the bias and wait for a few fs to allow the charge in the structure to rearrange before applying the pulse. Figure~\ref{fig3} shows the resulting particle current density $j(t)=-i[\Delta_{i}^{\perp}*G_{i}^{<}](t,t)$ for a given bias and photon energy. The inter-layer hybridization $\Delta_{i}^{\perp}$ follows from Eq.~(\ref{deltadef}) and the restriction to inter-layer hopping processes. Positive values of $j$ indicate triplons moving from the top to the bottom of the heterostructure and singlons in the opposite direction. After an initial backward flow induced by the change of polarization, the pulse produces charge carriers that will be eventually collected in the Ti leads. This current reflects the carrier multiplication described above: only high energy photons result in a carrier production after the pulse, yielding an increasing inter-layer current, as illustrated in the insets of Fig.~\ref{fig3}.

\begin{figure}
  \vspace*{6mm}
  \hspace*{-1mm}
  \includegraphics[width=0.228\textwidth]{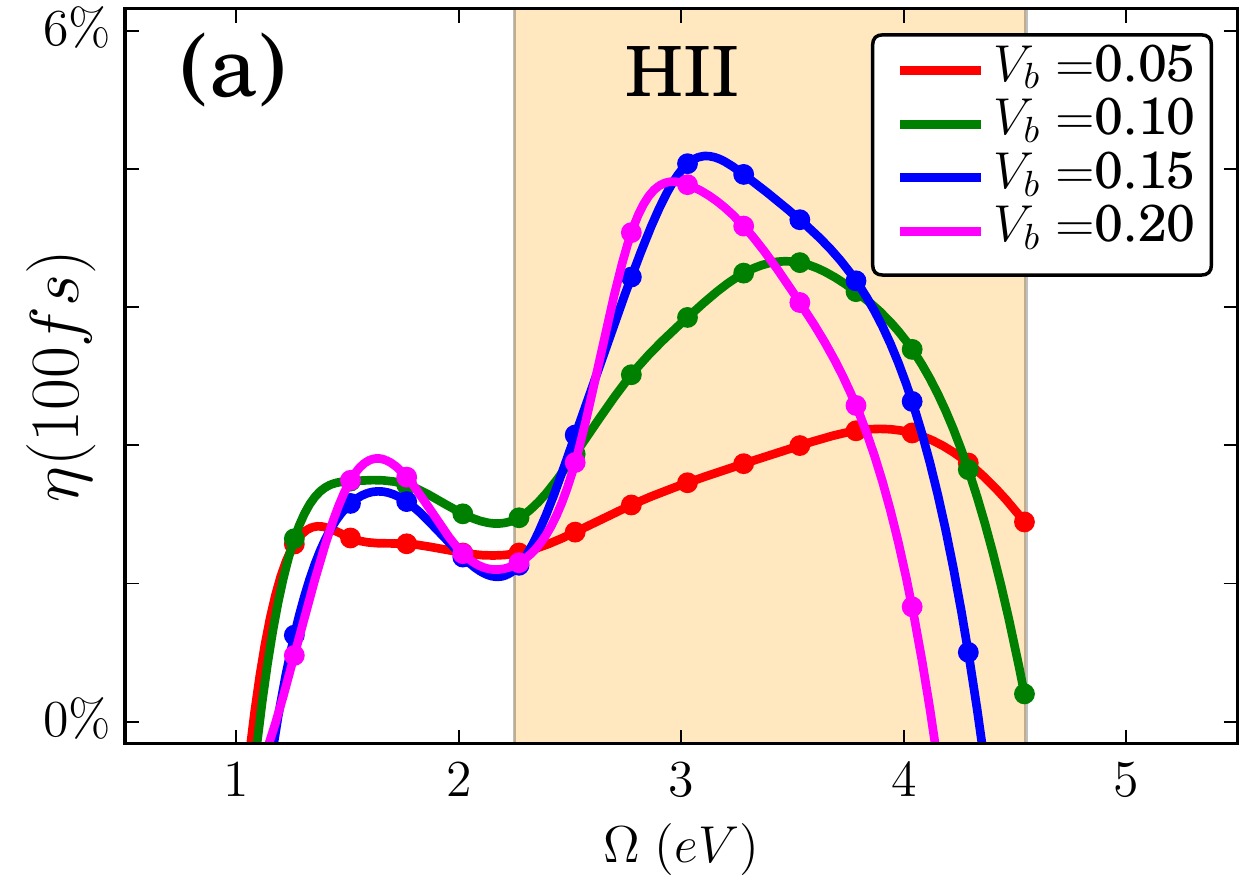}
  \hspace*{1mm}
  \includegraphics[width=0.228\textwidth]{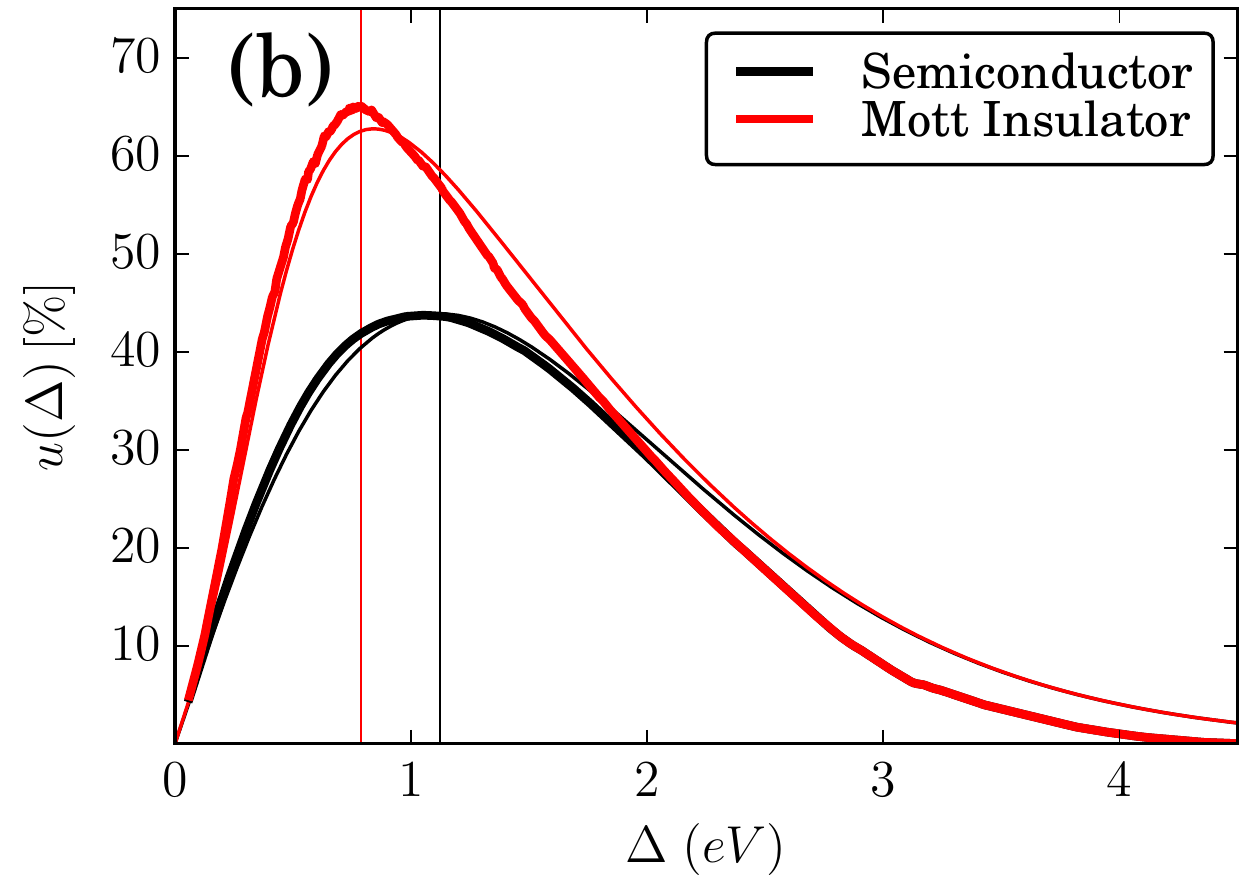}
  \caption{(a) Extractable energy fraction at $t=100$ fs for LVO. Pluse energies near $\Omega=3$ eV result in HII (see Fig.~\ref{fig2}d). (b) Ultimate efficiency of a semi-conductor solar cell (black) and Mott solar cell (red). The thin and thick lines show results for the black body and solar spectrum, respectively. \label{fig4}}
\end{figure}
The effect of impact ionization manifests itself in an increase of the current after high energy excitations on timescales which we can only reach by extrapolation. To better characterise the energy harvesting process, we define the energy ratio 
\begin{equation}
   \eta(t)=\frac{V_\text{bias}}{E_\text{abs}}\int_{0}^{t} \left[ j^{{\text{Ti}}^{\text{top}}}(t')+j^{{\text{Ti}}^{\text{bottom}}}(t')\right]dt',
   \label{eta}
\end{equation}
%
where $E_\text{abs}$ is the energy absorbed by the system right after the pulse, while $j^{{\text{Ti}}}$ is the current between the Ti layer and the adjacent bath. The integral represents the amount of charge collected by the leads. $\eta(t)$ thus measures the fraction of the absorbed energy that is converted into extractable potential energy at time $t$ for the given $V_\text{bias}$. The voltage bias affects the energy harvesting process in two ways: it increases the harvested energy per unit of charge and influences the charge separation process by counteracting the internal field \footnote{It also affects the polarization current at short times, but this is not relevant for the characterization of a real device.}.  Increasing $V_\text{bias}$ helps open transport channels by decreasing the DoS mismatch between neighboring layers and preventing high-energy carriers from being localized by the strong internal field \cite{Eckstein2014}. Since at the longest time that we can reach the inter-layer currents are still influenced by the (negative) polarization current, we extrapolate the harvested energy fraction to $100$ fs using the exponential fit $\eta_\text{fit}(t)=A \left[1-\exp  \left(t_0-t\right)/\tau\right]$. In Fig.~\ref{fig4}a we report the result for the LVO heterostructure. For all the reported biases it shows a comparable peak at low energy corresponding to the first charge excitation, which is always overlapping with the UHBs of the neighbouring layers. At higher energy $\eta$ increases with increasing $V_\text{bias}$: this behavior is not a consequence of a bigger absorption, since we normalize by $E_\text{abs}$, but rather of the increase in the carrier concentration induced by impact ionization and the improved alignment of the adjacent UHBs. An increase of $V_\text{bias}$ beyond 0.15 is detrimental to the harvesting process as the peak in $\eta$ shrinks to a narrower energy window, and eventually disappears at the breakdown voltage of the device. 
Fig.~\ref{fig4}a demonstrates that impact ionization plays an important role in converting high energy photons into additional harvestable charge carriers, thus contributing to a potentially high efficiency of Mott solar cells. 

In the LVO/STO structure, HII becomes relevant for photon energies in the tail of the solar light spectrum \cite{Assmann2013}. The optimal gap size for Mott solar cells is smaller than $1.1$ eV, since impact ionization requires pulse energies larger than twice the gap. The Shockley-Queisser (SQ) estimate for the ultimate efficiency $u[\Delta]$ of semi-conductor solar cells assumes that all charge excitations across a gap of size $\Delta$ contribute only an energy $\Delta$. In the presence of impact ionization, a simple generalization of the SQ argument yields
\begin{equation}
  u\left[\Delta\right]=\frac{\Delta\left[\sum_{i=1}^{M}i\int_{i\Delta}^{(i+1)\Delta}d\omega N(\omega)+\int_{(M+1)\Delta}^{\infty}d\omega N(\omega)\right]}{\int_{0}^{\infty}d\omega N(\omega)\omega} , 
  \label{effMott}
  \nonumber
\end{equation}
where we assume $M=3$ available excitations in the charge sector. In Fig.~\ref{fig4}b we compare the origial and Mott versions of the ultimate efficiency for photon densities $N(\omega)$ corresponding to black body radiation at 6000K and sunlight. The optimal gap size for Mott solar cells is approximately $0.8$ eV. A TMO Mott insulating compound characterised by a similar gap is YTiO$_3$ (YTO) \cite{Okimoto1995,Loa2007}. In the supplementary material, we provide simulation data for a YTO Mott solar cell and compare the results to the LVO/STO system. Some characteristics are similar, but the YTO results depend more sensitively on $V_\text{bias}$ due to the smaller gap. Since most of the spectral weight in YTO is concentrated in the high-spin sector, HII plays a minor role. Still, because the main absorption peak is at lower energy, this compound may be of interest for solar cell applications.

\paragraph*{Conclusions - }
We presented a semi-realistic study of the device characteristics of LVO based Mott solar cells. A relevant insight is the importance of HII for the harvesting of high energy photons and the cooperative effect between HII and KII, which can  result in a significant increase in the density of charge carriers on the 10 fs timescale. This timescale is short compared to decay processes associated with phonon excitations, so that impact ionization should indeed contribute to the efficient harvesting of solar energy in Mott systems \cite{Manousakis2010}. 

\paragraph*{Acknowledgements - }
This work was supported by the Swiss National Science Foundation through NCCR MARVEL and the European Research Council through ERC Consolidator Grant 724103. 
The calculations were performed on the Beo04 cluster at the university of Fribourg and the Piz Daint cluster at the Swiss National Supercomputing Centre (CSCS), using a software library developed by M. Eckstein and H. Strand. FP thanks Adriano Amaricci for useful discussions on the ED calculations, which were performed with a code developed at SISSA \footnote{https://github.com/aamaricci/dmft-ed}.

\bibliographystyle{apsrev4-1}
\bibliography{biblio}

\clearpage
\onecolumngrid
\begin{center}
\textbf{\large Hund excitations and the efficiency of Mott solar cells \\ - \\ Supplemental Material}
\vspace*{8mm}
\end{center}
\twocolumngrid

\paragraph*{DFT setup -}
The DFT calculations are performed using the \textsc{Quantum~ESPRESSO} package \cite{Giannozzi_et_al:2009} with the Perdew, Burke, and Ernzerhof (PBE) parametrization of the generalized gradient approximation (GGA) \cite{Perdew/Burke/Ernzerhof:1996} to the exchange-correlation functional. The bulk material with space group symmetry $Pbnm$ contains four symmetry-equivalent transition metal sites within the primitive unit cell. For the heterostructures we fix the in-plane lattice parameters to the theoretical STO lattice constant, and stack the films with the long orthorhombic axis parallel to the [001] growth direction, similar to Ref.~\onlinecite{Assmann2013}. Thereby, a glide plane $b$ parallel to the $c$-axis preserves the in-plane symmetry of the two transition-metal sites, such that we need to treat only one effective impurity problem per layer. The $c$-component of the cell and all internal coordinates are fully relaxed until the force components are smaller than 1 mRy/$a_0$ ($a_0$: Bohr radius) in the heterostructures (0.1 mRy/$a_0$ in bulk) and all the components of the stress tensor are smaller than 0.5 kbar (0.1 kbar). We employ scalar-relativistic ultrasoft pseudopotentials with the following semicore states included in the valence: 3s, 3p for V and Ti, 4s, 4p for Sr, and 5s, 5p for La. A plane-wave energy cutoff of 70 Ry for the wavefunctions and 840 Ry for the charge density is used, along with a 6 x 6 x 4 (bulk) or 8 x 8 x 4 (heterostructure) Monkhorst-Pack $k$-point grid. We use the Methfessel-Paxton smearing with a smearing parameter of 0.02 Ry for the broadening of the electron occupations for atomic relaxations, and 0.01 Ry for calculating the bandstructucture. The construction of the low-energy tight-binding Hamiltonian is performed using the \textsc{Wannier90} code \cite{Mostofi_et_al:2008}. Three MLWFs per site are constructed from inital projections onto atomic $t_{2g}$ orbitals. Since the energy window contains only the $t_{2g}$-derived bands around the Fermi level, the resulting MLWFs exhibit $p$-like tails on the surrounding O atoms. In order to obtain a non-interacting $\mathcal{H}_{0}$ (without local splittings due to spin and orbital order) as input for the DMFT calculations the Wannier functions are constructed from a non-spin-polarized calculation. To estimate the potential gradient in the insulating (spin and orbitally ordered) system, we use spin-polarized GGA+U calculations with $U_{\text{V}}=3.0$ eV and $U_{\text{Ti}}=9.8$ and C-type antiferromagnetic order. In this case the in-plane lattice constants are set to the average of those of bulk LVO with GGA+U.
\begin{figure}
  \includegraphics[trim=0 0 0 0 ,width=0.234\textwidth]{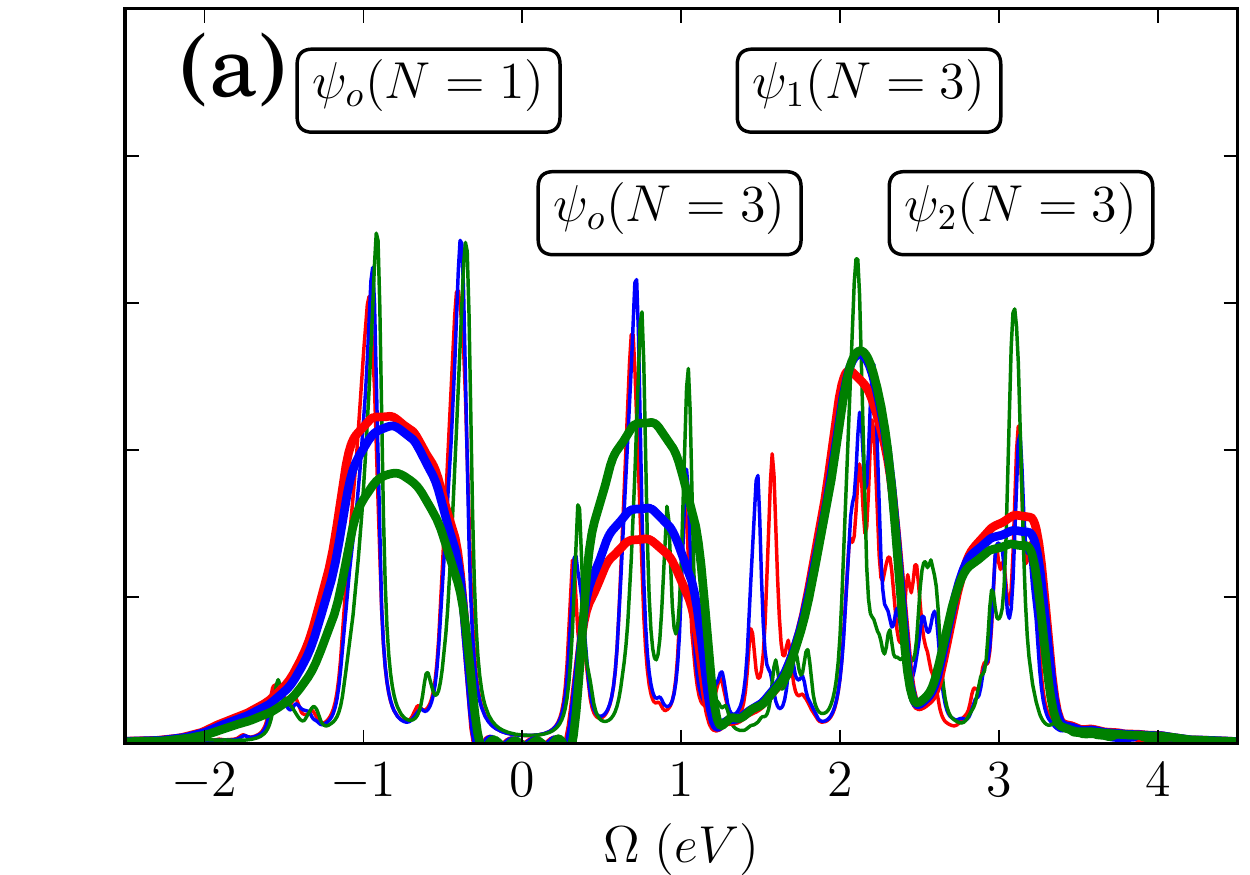}
  \includegraphics[trim=0 0 0 0 ,width=0.234\textwidth]{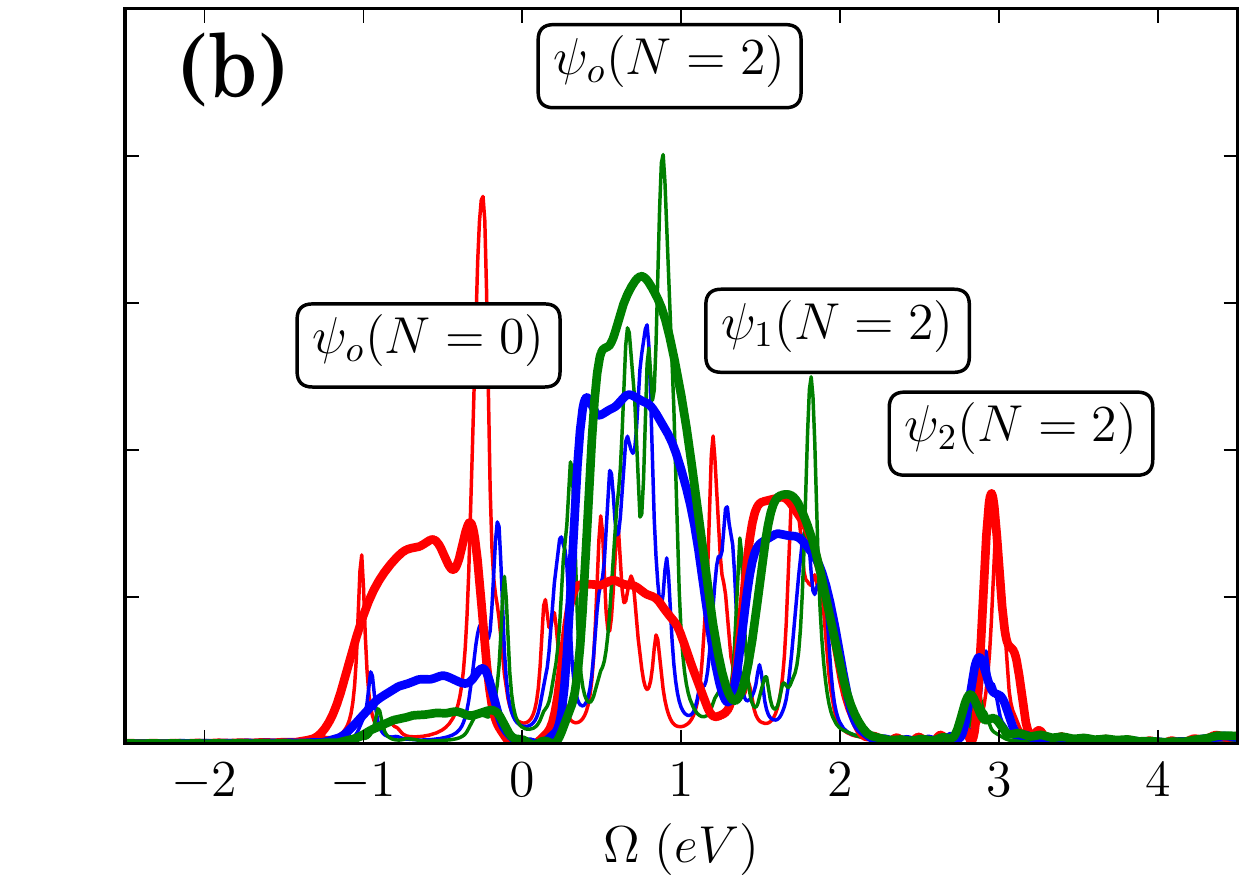}
  \includegraphics[trim=0 0 0 -8mm ,width=0.234\textwidth]{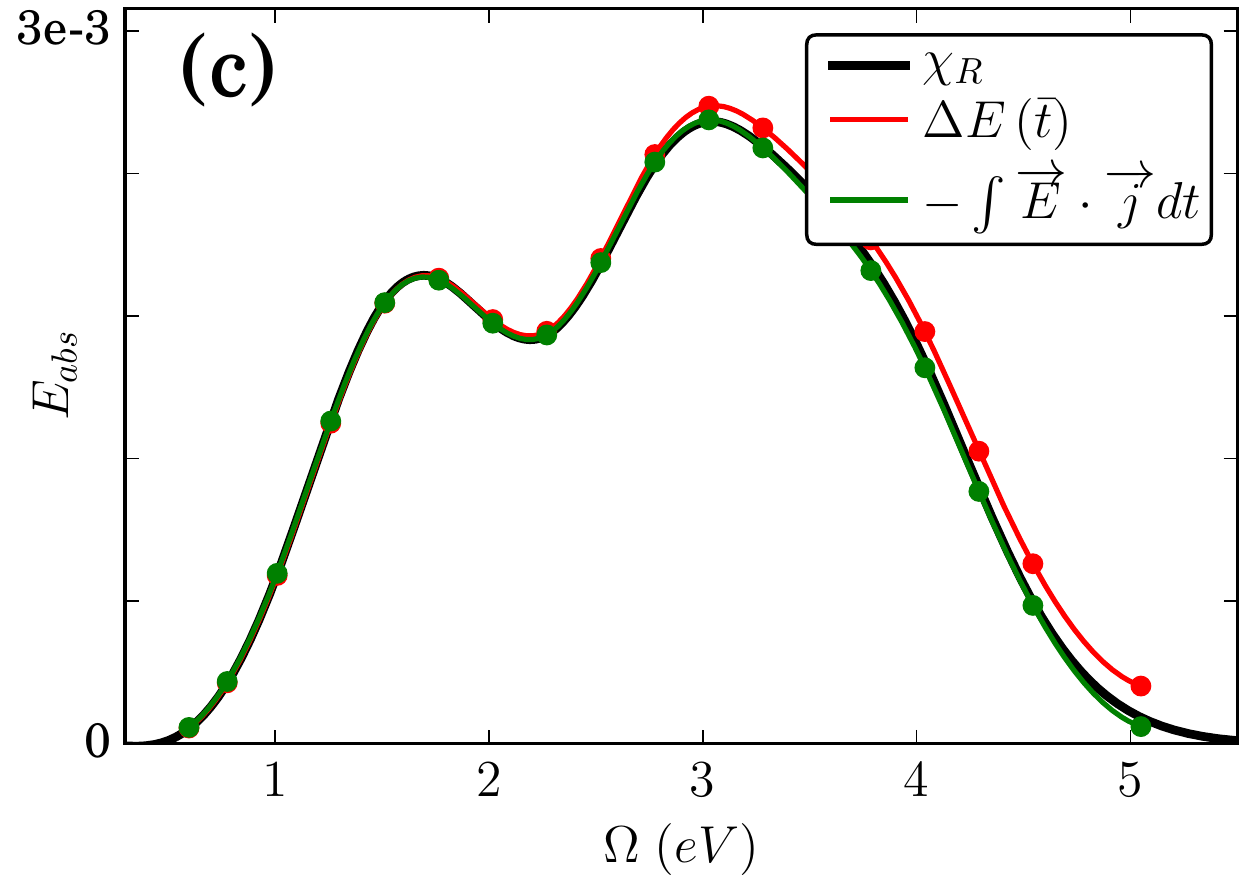}
  \includegraphics[trim=0 0 0 -8mm ,width=0.234\textwidth]{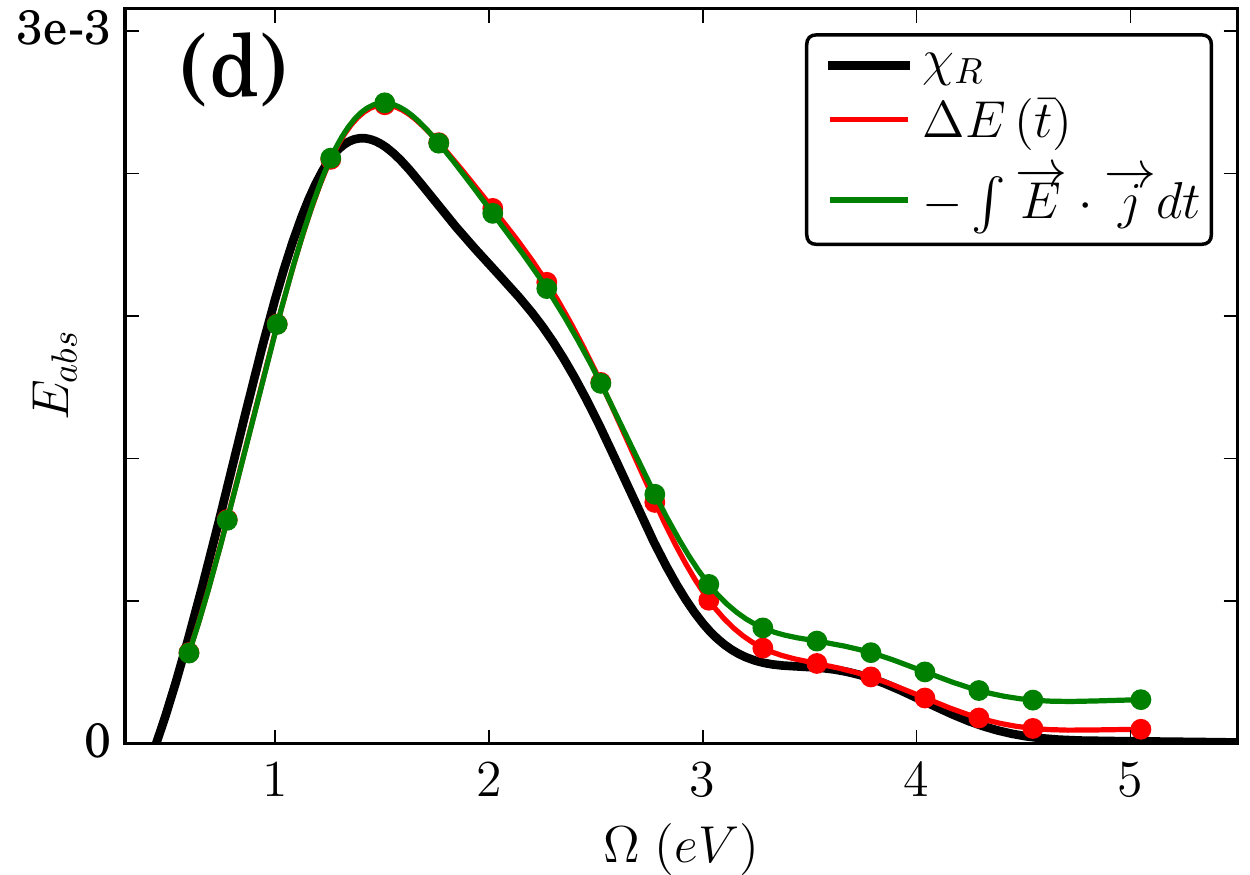}
  \includegraphics[trim=0 0 0 -8mm ,width=0.234\textwidth]{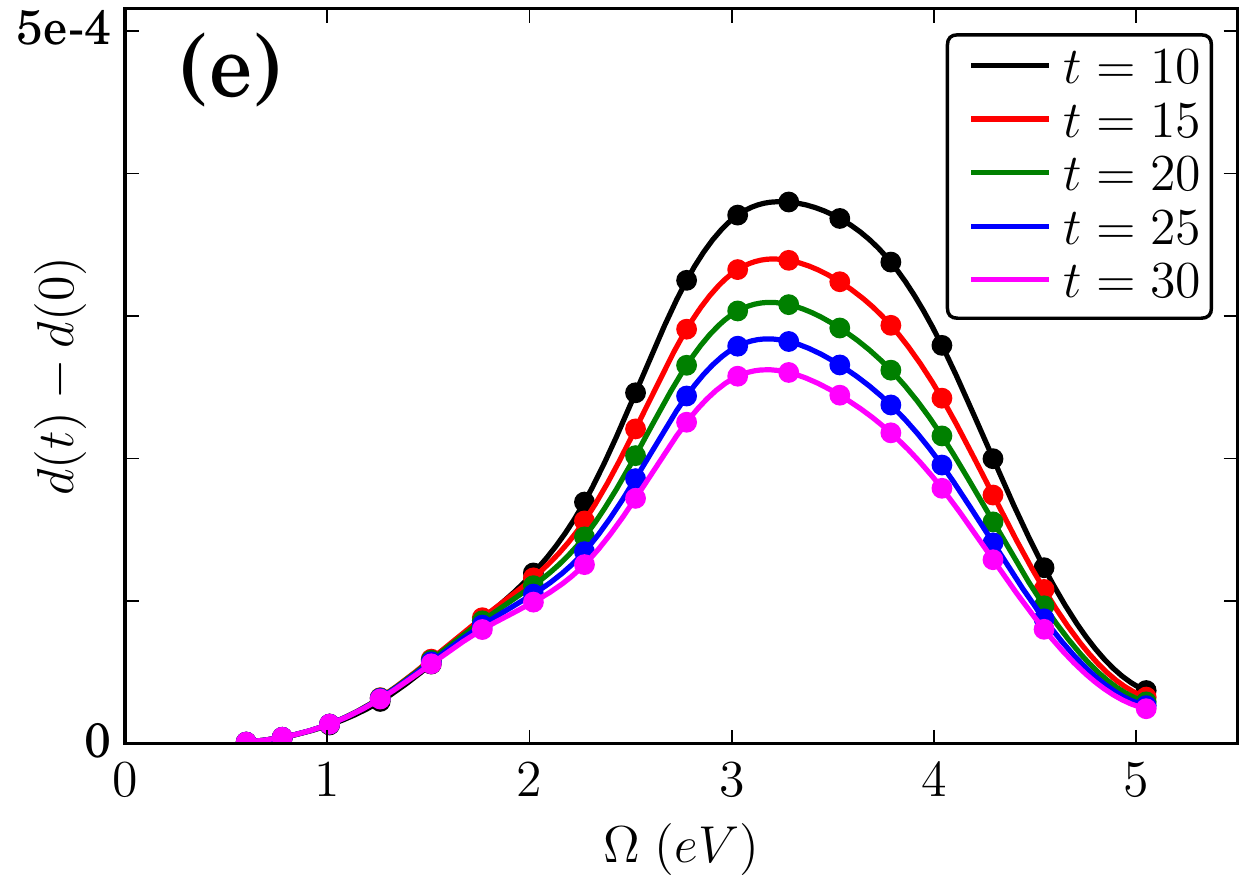}
  \includegraphics[trim=0 0 0 -8mm ,width=0.234\textwidth]{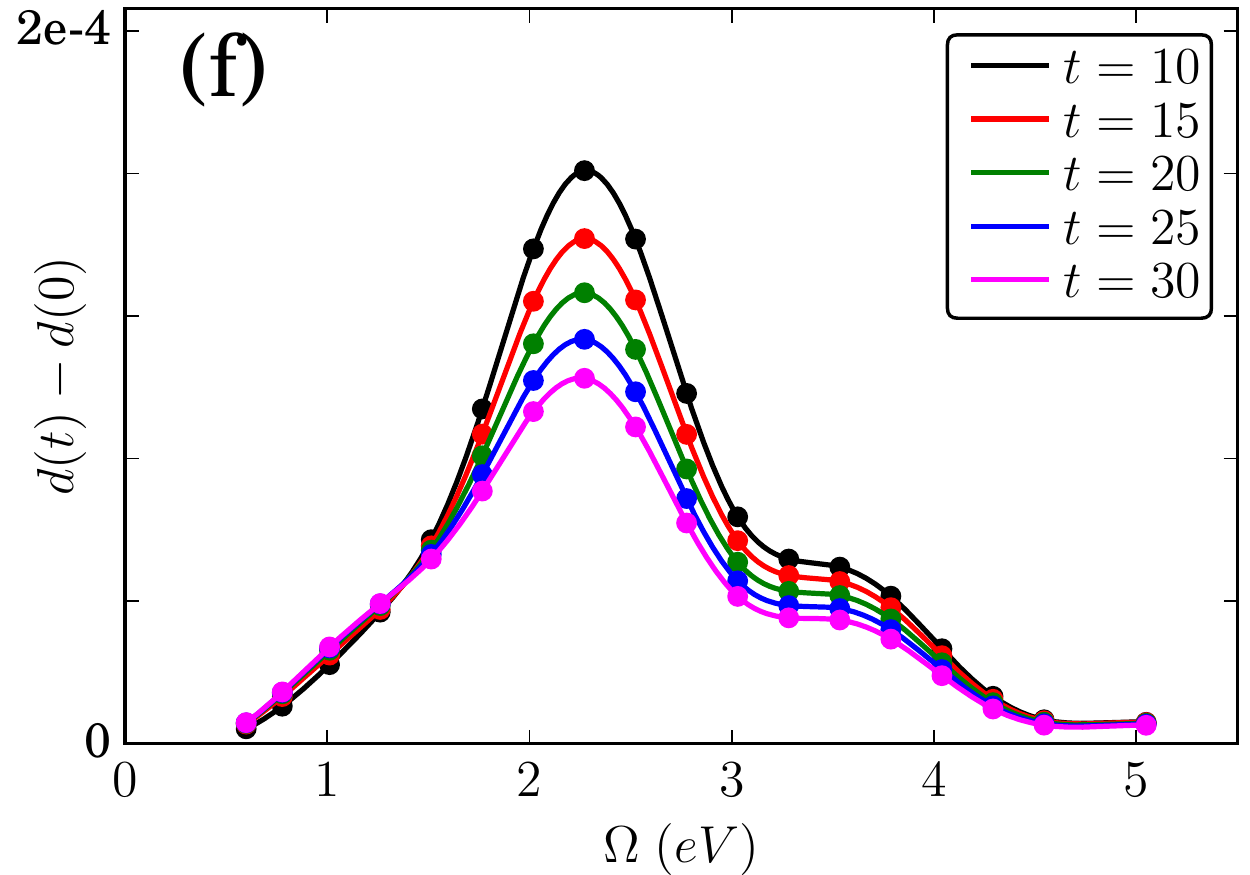}
  \caption{ (a,b) spectral function obtained with the NCA solver and approximate self-consistency of LVO and YTO respectively. (c,d) Absorption spectra of LVO and YTO computed with different methods: (black) weighted convolution between Hubbard bands, (red) energy variation after the pulse, (green) integral of absorbed power. (e,f) Double occupancy variation at different times. \label{fig1}}
\end{figure}
\paragraph*{Calculations for YTO -} In the DMFT calculations with the full lattice self-consistency and ED impurity solver, we fixed the Hund coupling to 0.64 eV and searched for the $U$ parameter that best reproduces the experimental spectral gap of the compound. For YTO, with a gap of $0.8$ eV, this procedure yields $U=3.5$ eV (compared to $U=4.5$ eV for LVO with a gap of $1.1$ eV). However, with this interaction parameter, the equilibrium calculation with the simplified self-consistency and NCA solver yields a metallic solution, forcing us to increase the interaction strength to $U=3.75$ eV, which results in the spectral functions of Fig.~\ref{fig1}b. In the case of LVO, the approximate treatment for $U=4.5$ eV yields a good agreement with the ED result (see Fig.~\ref{fig1}a), which indicates, as expected, that the approximate self-consistency and NCA are more reliable in more strongly correlated systems.  

The same parameters as used for the two bulk compounds have then been employed in the heterostructure setups. We report in Fig.~\ref{fig2} the separation rate extrapolated to $t = 100$ fs for the YTO/STO heterostructure. In this setup, some features of the pulse energy and voltage dependent $\eta$ are similar to the LVO/STO case (see Fig.~4a in the main text), in particular the appearance of inter-layer transport channels and the shrinking of the $\eta$ bell due to leakage currents. However the YTO/STO results depend more sensitively on $V_\text{bias}$ due to the smaller gap. Furthermore, in YTO the spectral weight of the UHB is concentrated on the low energy high-spin configuration peak, resulting in small HII effects (while KII is still effective). On the other hand, the main absorption peak in YTO is at lower energy than in LVO, which could make YTO more suited for solar cell applications. 
\begin{figure}
  \vspace{1cm}
  \includegraphics[width=0.3\textwidth]{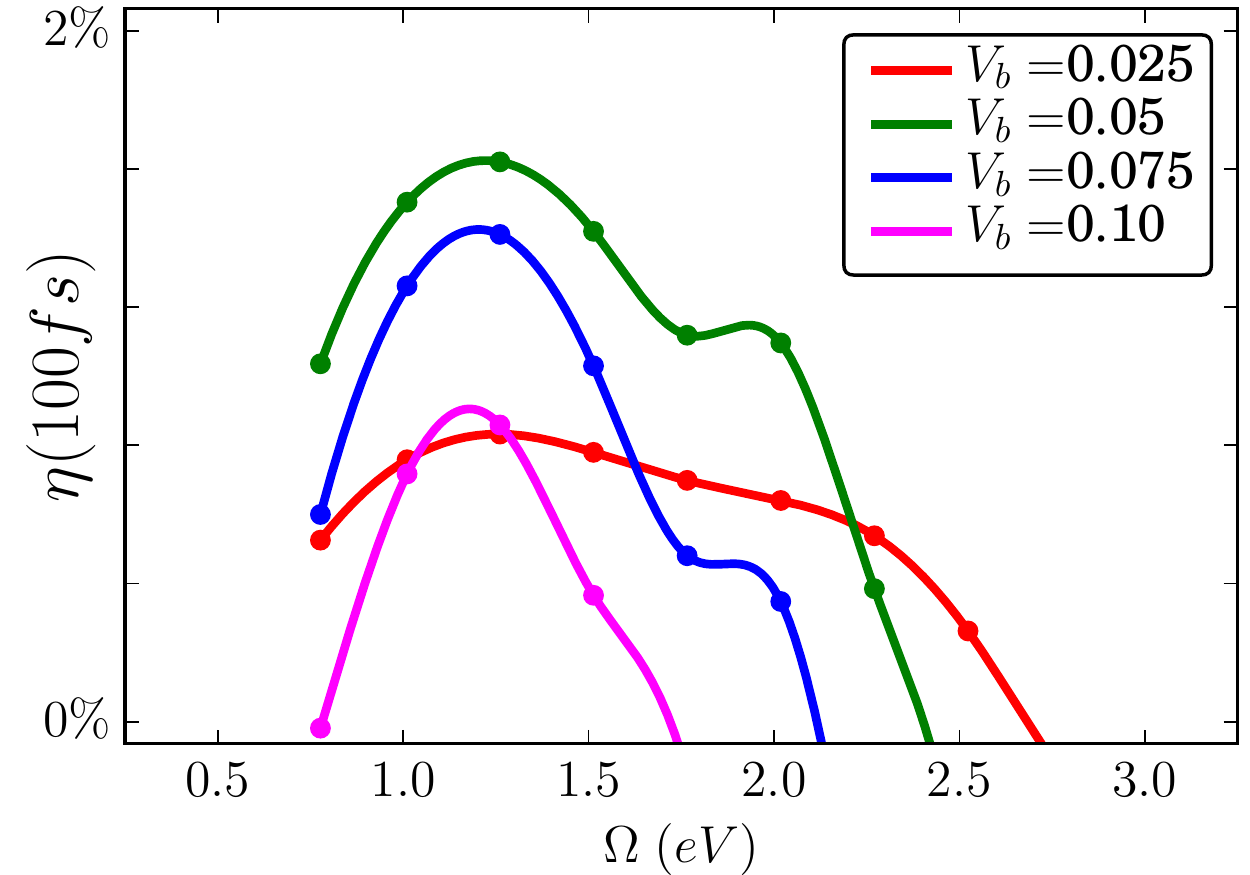}
  \caption{Separation rate extrapolated at $t = 100$ fs for the YTO/STO heterostructure as a function of pulse energy and applied bias. \label{fig2}}
\end{figure}

To investigate the effect of the spectral weight distribution on Impact Ionization we start by comparing in Fig.~\ref{fig1}(c,d) the bulk absorption spectra obtained with three different methods. i) Convolution between the lower Hubbard band (LHB) and the upper Hubbard band (UHB), weighted by the photon probability distribution: $\chi_{R}(\Omega)=\int d\omega f(\omega-\Omega) \int d\omega' A_\text{LHB}(\omega')A_\text{UHB}(\omega-\omega')$, where $f(\omega-\Omega)$ is a gaussian with fixed width and amplitude centered on the energy of the pulse $\Omega$. ii) Calculation of the total energy increase $\Delta E_\text{kin} + \Delta E_\text{pot}$ right after the pulse at $t=10$ fs. iii) Integral of the injected power. Apart from the gap amplitude, the main difference between the two compounds is that, in YTO, the spectral weight of the low-spin peaks ($|\psi_1(N=2)\rangle$ and $|\psi_2(N=2)\rangle$ in panel (b)), responsible for Hund Impact Ionization (HII), are much lower than in LVO, which explains why the carrier multiplication effects are weaker.
 
The information on the energy range in which HII is active can be deduced from Fig.~\ref{fig1}(e,f), where we plot the density of photo-induced double occupancies (doublons) at different time slices. Even though we excite two different ground states, with filling $N=2$ for LVO and $N=1$ for YTO, the particle-like excitations in the adjacent charge sector contain three families of states in both cases. The states with the lowest energy are high-spin configurations that in both cases do not contain any double occupancy, while the families of states at higher energy host low-spin configurations that contain doublons. Any process associated with HII must then be accompanied by a depletion of the doublon density happening on the same timescale as the carrier multiplication. Comparing Fig.~\ref{fig1}(e,f) with Fig.~\ref{fig1}(c,d) one sees that the doublon depletion occurs in the energy range compatible with a population of the two high-energy families of states in both systems. Figure~\ref{fig1}(e,f) demonstrates the presence of a low-to-high spin decay in YTO but with a considerably smaller magnitude than in the case of LVO (note the different $y$-axis scales). In particular, the process involving the low-spin state with the highest energy, which is the one producing the largest carrier multiplication effect in HII, is almost absent.  Again this is a consequence of the position and width of the peaks within the UHB, proving that the gap amplitude is not the only relevant parameter that needs to be tailored to take advantage of impact ionization. 
Our results point to an important role of the substructures of the Hubbard bands which result from local atomic physics. A lower Hund coupling would shift the Hubbard subbands associated with low-spin configurations to lower energies, which would be beneficial for light-harvesting purposes. However, $5J_\text{H}$ needs to be large compared to typical phonon energies, to avoid competition from phonon relaxation.

\end{document}